  \documentclass[prl,aps,twocolumn,showpacs]{revtex4}
\usepackage[dvips]{graphicx}
\usepackage{dcolumn}
\newcommand{\be}{\begin{equation}}
\newcommand{\ee}{\end{equation}}
\newcommand{\bea}{\begin{eqnarray}}
\newcommand{\eea}{\end{eqnarray}}
\newcommand{\nn}{ \nonumber}

\begin{document}
\topmargin=-20mm

\title{ On the de Haas -- van Alphen oscillations in quasi-two-dimensional metals: effect of the Fermi surface curvature}

\author{Natalya A. Zimbovskaya  }

\affiliation
{Department of Physics and Electronics, University of Puerto Rico at Humacao, Humacao, PR 00791 }

\begin{abstract}
 Here, we present the results of theoretical analysis of the de Haas-van Alphen oscillations in quasi-two-dimensional normal metals. We had been studying  effects of the Fermi surface (FS) shape on these oscillations. It was shown that the effects could be revealed and well pronounced when the FS curvature becomes zero at cross-sections with extremal cross-sectional areas. In this case both shape and amplitude of the oscillations could be significantly changed. Also, we analyze the effect of the FS local geometry on the angular dependencies of the oscillation amplitudes when the magnetic field is tilted away from the FS symmetry axis by the angle $\theta.$ We show that a peak appears at $\theta \approx 0$ whose height could be of the same order as the maximum at the Yamaji angle. This peak emerges when the FS includes zero curvature cross-sections of extremal areas. Such maximum was observed in experiments on the $\alpha-(BETS)_4TIHg(SeCN)_4.$ The obtained results could be applied to organic metals and other quasi-two-dimensional compounds.
 \end{abstract}

\date{\today}
\maketitle

\section{I. Introduction}

 Magnetic quantum oscillations are well known as a powerful tool repeatedly used in studies of electronic properties of various conventional metals \cite{1}. 
Theory of quantum oscillatory phenomena such as de Haas-van Alphen oscillations in the magnetization and Shubnikov-de Haas oscillations in the magnetoresistivity of conventional three-dimensional metals, was derived by Lifshitz and Kosevich (LK) in their well-known work \cite{2}. This theory was succesfully
employed to extract valuable informations concerning electron band-structure parameters from experimentally measured magnetic quantum oscillations.

In the last two decades quasi-two-dimensional (Q2D) materials with metallic-type conductivity (e.g. organic metals, intercalated compounds and some other) have attracted a substantial interest, and extensive efforts were applied to study their electron characteristics. Magnetic quantum oscillations are frequently used as a tool in these studies \cite{3,4}. A theory of magnetic oscillations in Q2D materials was proposed in several works (see e.g. Refs. \cite{5,6,7,8,9,10,11,12}). Significant progress is already made in developing the theory but there still remain some significant points not taken into account so far. The purpose of the present work is to contribute to the theory of de Haas-van Alphen oscillations in Q2D conductors by analyzing one of these points, namely, the effect of the Fermi surface (FS) curvature on the amplitude and shape of the oscillations.

It is already shown that local geometrical features of the FS may strongly affect quantum oscillations of elastic constants in both conventional and Q2D metals \cite{13,14}, so one has grounds to expect similar manifestations in the magnetization oscillations. The effect of the FS local geometry on  quantum oscillations in various observables could be easily given an explanation. In general, quantum oscillations are specified with contributions from vicinities of effective cross-sections of the FS. Those are cross-sections with minimum and maximum sectional areas. When the FS curvature becomes zero at an effective cross-section, the number of electrons associated with the latter increases, and their response enhances. This may significantly strengthen the oscillations originating from such cross-section and change their shape and phase.

\section{II. The model}

The FSs of Q2D metals are known to be systems of weakly rippled cylinders. Accordingly, the current theory adopts the tight-binding approximation to describe the energy-momentum relation for the charge carriers. So, the energy spectrum  could be written out as follows:
  \be 
 E {\bf (p)} = \sum_{n=0}  E_n (p_x,p_y) \cos \left(\frac{np_zd}{\hbar}\right).
  \ee
  Here, $z $ axis is assumed to be perpendicular to the conducting layers; $ p_x,p_y,p_z $ are the quasimomentum $ \bf p $ coordinates, and $ d $ is the interlayer distance. The dependence of the energies $ E_n $ of $ p_x, p_y $ is introduced in the Eq. (1) to take into account the anisotropies of the energy spectrum in the conducting layers planes. Due to the weakness of the interlayer interactions in Q2D conductors, the coefficient $ E_0 (p_x, p_y)$ is much greater than the remaining coefficients in the Fourier series in the Eq. (1). Usually, in studies of magnetic quantum oscillations in Q2D conductors the general energy-momentum relation (1) is simplified. The anisotropy of the energy spectrum in the layers planes is neglected, and only the first two perm in the sum over ``n" are kept. Then we arrive at the following simple model for a Q2D Fermi surface:
   \be 
 E{\bf (p)} = \frac{{\bf p}_\perp^2}{2 m_\perp} - 2t \cos \left (\frac{p_z d}{\hbar} \right).
   \ee
 where $ {\bf p_\perp} $ is the quasimomentum projection on the layer plane, and  $ m_\perp $ is the effective mass corresponding to the motion of quasiparticles in this plane.  The parameter $ t $ in this expression (2)  is the interlayer transfer integral whose value determines how much the FS is warped. When $ t $ goes to zero the FS becomes perfectly cylindrical.

Within this commonly used approximation the FS profile is completely established. When a strong magnetic field $ \bf B $ is applied along the normal to the layers $ \big ({\bf B} = (0,0,B)\big),$ the FS Gaussian curvature at the effective cross-sections $ K_{ex}$ is given by: 
  \be 
  K_{ex} = -\frac{1}{2 A_{ex}} \left (\frac{d^2 A}{dp_z^2} \right) = \pm \frac{2 \pi t m_\perp}{A_{ex}} \left(\frac{d}{\hbar} \right)^2
  \ee
  where $(d^2 A/dp_z^2)_{ex}$ means that the second derivative is calculated at the effective cross-section; $ A_{ex} $ is the cross-sectional area. So, the curvature takes on values proportional to those of the interlayer transfer integral $ t ,$ and it becomes zero only when the latter turns zero, and the FS passes into a perfect unwarped cylinder.

\begin{figure}[t]
\begin{center}
\includegraphics[width=8.8cm,height=8.8cm]{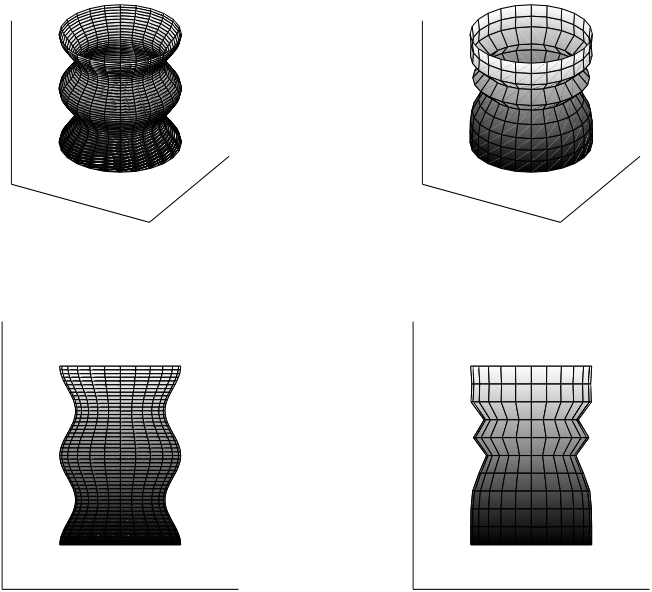}
\caption{Schematic plots of a Fermi surface with a cosine warping corresponding to the model (2) (left), and a Fermi surface of a complex profile including cross-sections with maximum areas where the FS curvature becomes zero (right).
}
\label{rateI}
\end{center}
\end{figure}

However, there are grounds to believe that the FSs of some realistic Q2D conductors possess more complex geometries than those described by the Eq. (2). For instance in the experiments on magnetic quantum oscillations in the layered perovskite oxide $Sr_2RuO_4$ \cite{15,16}, it was shown that to adequately describe the FS of this material one must take into consideration at least four terms in the expansion (1).  Here, we consider the Fermi cylinder corrugation of an arbitrary shape provided that the cylinder remains a surface of revolution. Separating out the first term in the expansion (1) we may rewrite the energy momentum relation in the form \cite{17}:
  \be 
  E {\bf (p)} = \frac{{\bf p}_\perp^2}{2m_\perp} - 2 t  \epsilon \left (\frac{p_zd}{\hbar} \right )
   \ee
  where
  \be 
\epsilon \left(\frac{p_z d}{\hbar}\right) = \sum_{n=1}^\infty \epsilon_n \cos \left(\frac{np_z d}{\hbar}\right)
  \ee
 with $ \epsilon_n = - E_n/2t.$ It follows from this expression that $ \epsilon (p_z d/\hbar) $ is an even periodic function of $ p_z $ whose period equals $ 2 \pi \hbar/d. $ Omitting all terms with $ n>1 $ and assuming $ E = 2t ,$ 
we may reduce our energy momentum relation (4) to the simple from (2). By introducing this expression we get opportunities to describe Q2D FSs of various profiles (see Fig. 1) and to analyze the influence of their fine geometrical features on the de Haas-van Alphen oscillations. As shown below, these studies bring some nontrivial results which could not be obtained within the simple approximation (2).

\section{III. Quantum oscillations in magnetization}

We start from the standard expression for the longitudinal magnetization:
  \be 
  M_{||} (B,T, \mu) \equiv M_z (B,T,\mu) = - \left ( \frac{\partial \Omega}{\partial B} \right )_{T,\mu}.
   \ee 
   Here, the magnetization depends on the temperature $ T $ and the chemical potential $ \mu. $ The latter itself is a function of the magnetic field and temperature, and oscillates in strong magnetic fields. The expression for the thermodynamic potential could be written out in a usual fashion:
   \be
   \Omega (B,T,\mu) = - T \sum \ln \Big \{1 + \exp \big[(\mu - E)/T\big] \Big\}
   \ee
   where the summation is  carried over all possible states of quasiparticles. When a strong magnetic field is applied the quasiparticles have the Landau energy spectrum, so the expression (4) takes the form:
   \be 
  E_{n,\sigma} (p_z) = \hbar \omega \left (n + \frac{1}{2} \right ) + \sigma  \frac{1}{2} g\hbar \omega_0 - 2 t \epsilon \left(\frac{p_z d}{\hbar} \right).
   \ee
  Here, $ \omega $ is the cyclotron frequency, $ \omega_0 = \beta B; \ \beta $ is the Bohr magneton, $ \sigma $ is the spin quantum number and $g$ is the spin splitting coefficient $(g$-factor). Accordingly, we rewrite Eq. (7) as follows:
   \bea 
  \Omega (B,T,\mu) &= &-\frac{T \omega}{(2 \pi \hbar)^2} \sum_{n,\sigma} \int_{-\pi\hbar/d}^{\pi \hbar/d} \ln \Big \{ 1  
  \nn\\ 
&&+ \exp \left[ \frac{\mu - E_{n,\sigma}(p_z)}{T} \right] \bigg \} m_\perp d p_z .
   \eea
  In further consideration we assume as usually, that the cyclotron quantum $ \hbar \omega $ is small compared to the chemical potential $ \mu. $ Then we employ the Poisson summation formula:
   \bea 
  \sum_{n=0}^\infty f \left(n + \frac{1}{2} \right) & =& \int_0^\infty f(x) \big [1 + 2 \mbox{Re} \sum_{r=1}^\infty (-1)^r 
   \nn \\ &&\times
\exp (2 \pi i r x) \big] dx.
   \eea
  Using this formula, the expression for the thermodynamic potential could be presented as a sum of a monotonous term $ \Omega_0 $ and an oscillating correction $ \Delta \Omega $:  
    \be 
  \Delta \Omega = \frac{i}{4 \pi^2 \hbar \lambda^2} \sum_\sigma \sum_{r=1}^\infty \frac{(-1)^r}{r} \int_0^\infty \frac{I (E_\sigma)dE}{1 + \exp[(E_\sigma - \mu)/T]} .
   \ee
  The function $ I(E_\sigma) $ is given by:
   \be 
  I (E_\sigma) = 2 i \int \exp \bigg [ ir \frac{\lambda^2}{\hbar^2} A (E_\sigma,p_z) \bigg] dp_z
   \ee
 where $ \lambda $ is the magnetic length, and $ A (E_\sigma, p_z) $ is the cross-sectional area.

Until this point we followed LK theory in derivation of the expression for $ \Delta \Omega. $ As a result we arrived at Eqs. (11),(12) which are valid for conventional 3D metals as well as for Q2D and perfectly 2D conductors. Diversities in the expressions for $ \Delta \Omega $ appear in the course of calculations of the function $ I (E_\sigma).$ These calculations  bring different results for different FS geometries. In deriving the standard LK formula it is assumed that the FS curvature is nonzero at the effective cross-sections with the extremal areas, and $ I(E_\sigma, p_z) $ is approximated using the stationary phase method. For 2D metals the calculations of $ I (E_\sigma) $ are trivial for the FS is a cylinder and the cross-sectional area $A$ does not depend of $ p_z. $ Obviously, in this case the FS curvature is everywhere zero. The oscillating part of the thermodynamic potential $ \Omega $ for a 2D metal takes on the form:
   \be 
  \Delta \Omega = N \hbar \omega \left ( \frac{B}{F} \right) \sum_{r=1}^\infty \frac{(-1)^r}{(\pi r)^2} R_T(r) R_S(r) \cos\left(\frac{2 \pi r F}{B} \right).
   \ee
  Here,  $ F= c A/2 \pi \hbar e;\ N $ is the  density of charge carriers, and $R_T(r), R_S (r) $ describe the effects of temperature, and spin splitting.  Also, the scattering of electrons deteriorates magnetic quantum oscillations for it causes energy levels broadening. The simplest way to account for the effects of electron scattering on the oscillation amplitudes is to introduce an extra damping factor $R_D (r) $ (Dingle factor) into the expression (13). The usual approximation for the latter has the form $ R_D (r) = \exp (- 2\pi r/\omega \tau) $ \cite{1} where $ \tau $ is the scattering time of electrons. In further calculations we adopt this simple form for $ R_D(r) $ for more sophisticated expressions are irrelevant to the main point of our subject. As a result we obtain:
  \bea 
 \Delta \Omega & =& N \hbar \omega \left(\frac{B}{F}\right)
\sum_{r=1}^\infty \frac{(-1)^r}{(\pi r)^2}
 \nn \\ &\times &
 R_T (r) R_S (r) R_D(r) \cos\left(\frac{2\pi rF}{B}\right)
 \nn \\ &\equiv& 
 N \hbar \omega \left(\frac{B}{F}\right) \sum_{r=1}^\infty 
\frac{(-1)^r}{(\pi r)^2} R (r)  \cos\left(\frac{2\pi rF}{B}\right).
  \eea
  Correspondingly, we arrive at the following result for the oscillating part of the longitudinal magnetization \cite{10}:
  \be
  \Delta M_{||} = - 2 N\beta \frac{\omega}{\omega_0} \sum_{r=1}^\infty\frac{(-1)^r}{\pi r}R(r) \sin\left(2\pi r\frac{F}{B}\right).
   \ee  
 Taking into account the $ p_z $ depending term in the charge carriers energy spectrum within the approximation (2)
one can expand the integrand in the Eq.(12) in Bessel functions and easily carry out integration over $ p_z.$ Then $ \Delta M_{||} $ takes the form \cite{18}:
   \be
  \Delta M_{||} = - 2 N\beta \frac{\omega}{\omega_0} \sum_{r=1}^\infty\frac{(-1)^r}{\pi r}R(r) J_0 \left(\frac{4\pi r t}{\hbar \omega}\right) \sin\left(2\pi r\frac{F}{B}\right).
   \ee  
  When the FS warping is negligible $(t\ll \hbar \omega) $ this expression passes into the previous formula (15)  describing the magnetization of a 2D metal. In the opposite limit $ (t > \hbar \omega) $ one can use the corresponding asymptotic for the Bessel functions.
As a result the expression for $ \Delta M_{||} $ is transformed to the form similar to the LK result for the conventional 3D metals:
   \bea 
 \Delta M_{||} &=& -2 N \beta \frac{\omega}{\omega_0}\sqrt{\frac{\hbar \omega}{2 \pi^2 t}} \sum_{r=1}^\infty \frac{(-1)^r R(r)}{\pi r^{3/2}}   
  \nn \\  &\times &
\sin \left (2 \pi r \frac{F}{B} \right ) 
\cos \left (\frac{4 \pi rt}{\hbar \omega} - \frac{\pi}{4} \right)
\nn\\
 &=& - N \beta \frac{\omega}{\omega_0} \sqrt{\frac{\hbar \omega}{2\pi^2 t}}
\sum_{r=1}^\infty \frac{(-1)^r R(r)}{\pi r^{3/2}}
  \nn\\ &\times& \!
\bigg\{\!\sin \left (\frac{2\pi r F_{\max}}{B} - \frac{\pi}{4} \right) 
 + \sin\left (\frac{2\pi r F_{\min}}{B} + \frac{\pi}{4} \right)\! \bigg\} \nn\\
  \eea
  where $ F_{\max} = F A_{\max}/A,\ F_{\min} = F A_{\min}/A,\ A_{\max}, A_{\min}$ are the maximum and minimum cross-sectional areas, respectively. As before, $A $ is the cross-sectional area of the incorrugated cylindrical FS.
  Before we proceed we remark again that the commonly used  approximation (2) is not suitable to analyze the effects of the FS shape in Q2D metals. Within this model, the resulting formulas for $ \Delta M $ are either reduced to the 2D limit \big[Eq. (15)\big] or they describe only cosine warped profiles of the FS without any variations \big[Eqs. (16), (17)\big]. In both cases some important features of the oscillations could be missed.

 We may expect the effect of the FS curvature on the magnetization oscillations to appeare when the FS warping is distinct $(t > \hbar\omega). $ To analyze these effects we return back to our generalized energy-momentum relation (4).  Then we assume that the FS curvature becomes zero at the effective cross-section at $ p_z = p^*. $ Then, as follows from the expression for the FS curvature (3), $d^2 A/dp_z^2 $ equals zero at $ p_z = p^*, $ so
 we can write the following approximation for the cross-sectional area:
   \be 
  A (p_z) \approx A_{ex} \pm \frac{1}{(2l)!} \left| \frac{d^{2l} A}{d p_z^{2l}} \right |_{p_z=p^*} (p_z-p_*)^{2l}
   \ee
  where $ l > 1. $

Performing integration in parts in the expression (11) we obtain
  \be 
 \Delta\Omega = \frac{\hbar \omega}{\pi^2\hbar\lambda^2}
\sum_{r=1}^\infty \frac{(-1)^2 }{(\pi r)^2} R(r)
\int_0^{\pi\hbar/d} \cos \left(\frac{r\lambda^2 }{\hbar^2}
A(p_z) \right) dp_z
  \ee
  Then, using the approximation (18), and applying the stationary phase method to compute the integral over $ p_z $ in the expression (19), we obtain the following result for the contribution $\delta\Omega $ from the nearly cylindrical cross-section to the oscillating part of $ \Omega: $
   \bea 
  \Delta \Omega &=& 2 N \alpha_l \left (\frac{B}{2F} \right)^\rho \hbar\omega \sum_{r=1}^\infty \frac{(-1)^r}{(\pi r)^{\rho+1}}R(r) 
 \nn\\ &&\times
\cos \left ( 2\pi r \frac{F_{ex}}{B} \pm \frac{\pi}{4 l} \right).
   \eea
  Here, $ \rho = 1 + 1/2l $ ;
      \be 
 \alpha_l = \frac{A_{ex}}{V_0} \Gamma (\rho) \left(
\frac{(2l)! A_{ex}}{2\pi m_\perp t\big|d^{2l}\epsilon/dp_z^{2l} \big|_{p_z=p^*}}\right)^{1/2l}
  \ee
  $ V_0 $ is the FS volume in a single Brillouin zone, and  $ \Gamma (\rho) $ is the gamma function. Basing on this expression (20) we  get the corresponding term in the oscillating part of the magnetization:
   \bea 
  \Delta M_{||}& =& - 2 N \alpha_l \beta \left (\frac{B}{F}\right)^{1/2l} \frac{\omega}{\omega_0}\sum_{r=1}^\infty \frac{(-1)^r}{(\pi r)^\rho} R(r)
\nn \\ &&\times 
\sin \left (2\pi r \frac{F_{ex}}{B}  \pm \frac{\pi}{4l} \right)
\nn \\
&\equiv& - 2N \xi_l \left(\frac{\hbar\omega}{t}\right)^{1/2l}
\sum_{r=1}^\infty \frac{(-1)^r}{(\pi r)^\rho} R(r) 
  \nn \\ && \times
\sin \left(2\pi r \frac{F_{ex}}{B} \pm \frac{\pi}{4l}\right).
     \eea
  where
 \be 
 \xi_l = \frac{A_{ex}}{V_0} \Gamma (\rho) \left(
\frac{(2l)! }{\big|d^{2l}\epsilon/dp_z^{2l} \big|_{p_z=p^*}}\right)^{1/2l}
  \ee
 To arrive at the complete expression for $ \Delta M_{||} $ we must sum up terms originating from all effective cross-sections of the FS.

  The FS shape near $ p_z = p^* $ is determined by the shape parameter $ l. $ When $ l=1 $ the FS has a nonzero curvature at the considered cross-section. In this case Eq. (22) agrees with the well-known LK result. Assuming that there are two extremal cross-sections (with the minimum and maximum cross-sectional areas, respectively), and summing up the contribution from the both we can easily transform our result (22) to the form (17).

When $ l\gg 1$ we may roughly estimate $\big|d^{2l}\epsilon/dp_z^{2l} \big|_{p_z=p^*} $ as $(2l)^{2l+1}/(2l +1)$ and $ (2l)! $ as  $\exp[(2l + 2)\ln(2l+2)] $ (see \cite{19}). So, $\lim_{l \to \infty} \xi_l = 1/2, $ and Eq. (22) passes into the expression (15) describing magnetization oscillations in 2D conductors multiplied by $1/2.$ This extra factor appears because the expression (22) describes the contribution from a single nearly cylindrical cross-section of the FS.  When the shape parameter for both effective cross-sections goes to infinity, their contributions to the magnetization oscillations become identical, and putting them together we arrive at the expression (15). In general, one may treat $ l $ as a phenomenological parameter whose actual value could be found from experiments. The greater is the value of this parameter the closer is the FS to a cylinder near $ p_z = p^*. $  

Oscillations in magnetization described by the expression (22) vary in magnitude, shape and phase depending on the value of the shape parameter $ l $ which determines the FS local geometry near the effective cross-section. This is illustrated in the Fig. 2. As shown in this figure, when there is the close proximity of the FS near $ p_z = p^* $ to a cylinder, the oscillations are sawtoothed and resemble those occuring in 2D metals \cite{7,11} or originating from cylindrical segments of the FSs in conventional 3D metals \cite{20,21}. When the FS curvature takes on nonzero value at $ p_z = p^* \ (l=1)$ the  oscillations are similar to those in conventional metals. Here, we emphasize the difference between our result (22) and the expression (16). Using the latter one could easily obtain sawtoothed oscillations typical for 2D metals but only for small values of the transfer integral $ t \ (t \ll \hbar \omega) $ when the FS crimping is negligible. The present result (22) shows that the oscillation shape and phase may be determined not by the value of $ t $ itself but rather by the form of the function $ \epsilon (p_z d/\hbar) $ specifying the FS profile. The sawtoothed oscillations in magnetization could occur at $ t \sim \hbar \omega , $ when the FS curvature becomes zero at an effective cross-section. To ease the interpretation of this point one may imagine a FS shaped as a step-like cylinder. The curvature of such FS is everywhere zero, and oscillations from both kinds of the cross-sections (with minimum and maximum cross-sectional areas, respectively) should be similar to those in 2D metals. Nevertheless, the difference in the cross-sectional areas (the FS crimping) could be well pronounced, and a beat effect could be manifested. Obviously, this effect is absent when $ t \ll \hbar\omega $ and the FS warping is negligible. 

\begin{figure}[t]
\begin{center}
\includegraphics[width=8.8cm,height=4.4cm]{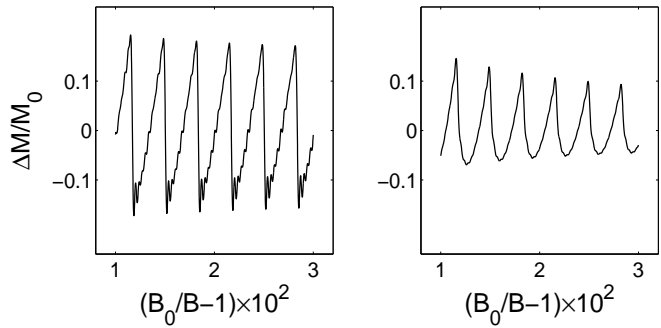}
\caption{De Haas-van Alphen oscillations described by the Eq. (22) for $ l=4 $ (left panel) and $ l=1 $ (right panel). Calculations are carried out for $ T = T_D = 0.5 K, \ B_0 = 10 T, \ F/B_0 = 300; \ T_D $ is the Dingle temperature, and $ M_0 = 2 N \beta \omega/\omega_0. $
}
\label{rateI}
\end{center}
\end{figure}
             
Also, it may happen that the FS curvature becomes zero at some effective cross-sections and remains nonzero at the rest of them. Then the contributions from zero curvature cross-sections $ (l>1) $ would exceed in magnitude those originating from the oridinary cross-sections $(l=1). $ This follows from the expression (22) where the factor $ (\hbar\omega/t)^{1/2l}\ (\hbar\omega < 1) $ is included. Depending on the value of the shape parameter $ l $ this factor takes on values between $(\hbar\omega/t)^{1/2}\ (l=1)$
and $1\ (l \to \infty). $ So, when there is a close proximity of the FS to a pure cylinder at some extremal cross-sections, the contributions from these cross-sections would be predominating, and they will determine the shape and amplitude of the magnetization oscillations in whole.

\section{IV. The effect of the chemical potential oscillations}

It is known that chemical potential $ \mu $ oscillates in strong (quantizing) magnetic fields. These oscillations and the oscillations in the magnetization are closely related and described by similar series. The chemical potential is determined by the equation:
   \be 
  N = \left (\frac{\partial \Omega}{\partial \mu} \right)_{T,B}. 
  \ee
  When the quantizing magnetic field is applied, the charge carriers density gets an oscillating correction $ \Delta N (\mu) :$
  \be 
 N = N_0 (\mu) + \Delta N (\mu)
  \ee
 where $ N_0 $ is the charge carriers density at $ B= 0. $ Provided that the charge carriers density is fixed, this correction is to be balanced by an oscillating term $\Delta \mu$ which appears in the chemical potencial due to the magnetic field. These corrections are related to each other by the equation \cite{1}:
  \be 
 \Delta\mu \frac{\partial N_0}{\partial \mu} = \Delta N (\mu)
  \ee
  where $ \partial N_0/\partial \mu \equiv D_0 $ is the charge carriers density of states at the FS in the absence of the magnetic field. So, we have:
  \be 
  \Delta \mu (B) = \frac{1}{D_0} \left(\frac{\partial \Delta \Omega}{\partial \mu}\right)_{T,B}
  \ee
  Assuming that the FS curvature becomes zero at the effective cross-section at $ p_z = p^*, $  using the expression (20) for $ \Delta \Omega, $ and omitting smaller contributions from the remaining cross-sections of nonzero curvature we obtain:
   \be 
  \Delta \mu = - \zeta_l \hbar \omega \sum_{r=1}^\infty \frac{(-1)^r}{(\pi r)^\rho} R(r) \sin\left (2 \pi r\frac{F_{ex}}{B} \pm \frac{\pi}{4l} \right).
   \ee
  Here, $ \zeta_l = 4N\alpha_l (B/F)^\rho/\hbar \omega D_0, $ and $ D_0 $ is the electron density of states at the FS in the absence of the magnetic field. The dimensionless factor $\zeta_l $ takes on small values of the order of $(\hbar \omega/E_F)^{1/2l}.$  This result (28) as well as the expression for the oscillating part of magnetization (22) do not contradict the corresponding results for 2D and 3D metals reported in earlier works \cite{10,22}. For a conventional 3D metal whose FS has a nonzero curvature at the effective cross-sections the oscillating correction $ \Delta \mu $ is very small compared to $E_F \ \big (\Delta\mu/E_F \sim (\hbar\omega/E_F)^{3/2}\big).$ Therefore it does not bring noticeable changes in de Haas-van Alphen oscillations. However, in the case of Q2D metals this correction could be more important \cite{10,11}. To analyze the effect of the quantum oscillations in chemical potential, one must take into account that the extremal cross-sectional areas $ A_{ex} $ include corrections proportional to $ \Delta \mu, $ namely:
  \be 
  A_{ex} = A_{ex} (0) + 2\pi m_\perp \Delta \mu (B)
    \ee
  where $ A_{ex} (0) $ is the cross-sectional area in the absence of the magnetic field. Correspondingly, the argument of the cosine function in the Eq. (20) must be written in the form:
   \be 
  2 \pi r \frac{F_{ex}}{B} \pm \frac{\pi}{4l} = 2 \pi r \left (\frac{F_0^{ex}}{B} + \frac{\Delta \mu}{\hbar \omega}\right ) \pm \frac{\pi}{4l}.
   \ee
  To simplify further analysis we keep only the first term in the expansion (28). Then we can employ the identity:
   \bea 
  \exp \left [ ir \zeta_l R(1) \sin \left(2\pi \frac{F_0^{ex}}{B} \pm \frac{\pi}{4l} \right ) \right ]
   \nn \\  =
  \sum_{n=-\infty}^\infty J_n \big(r \zeta_l R(1)\big) \exp \left [in \left(\frac{2 \pi F_0^{ex}}{B} \pm \frac{\pi}{4l} \right) \right ]
   \eea
  where $ J_n \big(r \zeta_l R(1)\big) $ are the Bessel functions. 

Using this identity we can write the expression for $ \Delta M_{||} $ as follows:
   \bea 
   \Delta M_{||} &= &- 2N \alpha_l \beta \left (\frac{B}{F}\right )^{1/2l}\frac{\omega}{\omega_0}
  \nn \\ && \times
 \sum_{s=1}^\infty \bigg \{\!A_s \sin \left (2\pi s \frac{F_0^{ex}}{B} \pm \frac{\pi s}{4l} \right)
   \nn \\ && \mp
  B_s \cos\left(2 \pi s\frac{F_0^{ex}}{B} \pm \frac{\pi s}{4l} \right) \bigg \}.
  \eea
  Here, the coefficients $ A_s,B_s $ are given by:
  \bea 
  A_s &=& \sum_{r=1}^\infty \frac{(-1)^r R(r)}{(\pi r)^\rho} \bigg \{J_{r+s} \big(r \zeta_l R(1) \big ) \cos \left(\frac{\pi (r+1)}{4l} \right )
    \nn \\ 
  &&- J_{r-s} \big (r \zeta_l R(1) \big) \cos \left(\frac{\pi (r-1)}{4l} \right) \bigg \};
   \\ \nn\\
  B_s &=& \sum_{r=1}^\infty \frac{(-1)^r R(r)}{(\pi r)^\rho} \bigg \{J_{r+s} \big(r \zeta_l R(1) \big ) \sin \left(\frac{\pi (r+1)}{4l} \right )
    \nn \\ 
  &&- J_{r-s} \big (r \zeta_l R(1) \big) \sin \left(\frac{\pi (r-1)}{4l} \right) \bigg \}.
   \eea
  These formulas (32)--(34) are generalizations of the results obtained for 2D metals \cite{11,23,24}.

As follows from Eqs. (32)--(34) the oscillating correction to the chemical potential may bring some changes in the amplitude and shape of the de Haas-van Alphen oscillations in Q2D metals. Keeping in mind that the damping factor $ R(r) $ takes on values less than unity, and reduces while $ r $ increases, we can write explicit expressions for a few first harmonics  in the form:
   \bea 
  \delta M_1 &=& 2 N \alpha_l R(1) \pi^{-\rho} \sin \left(2 \pi \frac{F_0^{ex}}{B} \pm \frac{\pi}{4l} \right );
  \\ \nn \\      
 \delta M_2  &=& - 2 N \alpha_l R(2)  (2\pi)^{-\rho} \bigg \{ \sin \left(4\pi \frac{F_0^{ex}}{B} \pm \frac{\pi}{4l} \right)
   \nn \\  &&+
 \frac{R^2(1)}{R(2)} 2^{1/2l} \zeta_l \sin \left(4\pi \frac{F_0^{ex}}{B} \pm \frac{\pi}{2l} \right) \bigg \};
\\ \nn \\ 
 \delta M_3 & = & 2 N \alpha_l R(3)  (3\pi)^{-\rho} \bigg \{
\sin \left(6\pi \frac{F_0^{ex}}{B} \pm \frac{\pi}{4l} \right)
   \nn \\   && +
 \frac{R(2)R(1)}{R(3)} \left(\frac{3}{2}\right)^\rho \zeta_l \sin \left(6\pi \frac{F_0^{ex}}{B} \pm \frac{\pi}{2l} \right)
 \bigg \}. \nn \\
   \eea  
  So we see that the chemical potential oscillations do not affect the fundamental harmonic but they contribute to higher harmonics bringing changes to their amplitude and phase. Similar conclusions were recently made analyzing the effect of chemical potential oscillations on magnetization in 2D metals \cite{11}. The effect depends on the FS local geometry. When the shape parameter $ l $ takes on greater values, the effect becomes stronger. As shown in the Fig. 3, for a pronounced proximity of the FS to a cylinder near the effective cross-section $ (l=4), $ a noticeable difference in the magnetization oscillations shape and magnitude arises due to the effect of the chemical potential oscillations.

  The correction from $ \Delta \mu $ brings changes in the position of spin-splitting zeros in the harmonics of the de Haas-van Alphen oscillations. These changes were studied before within the simple approximation (2) for the energy spectrum \cite{26}. It was shown that the spin-zero positions for the second and third harmonics are not completely determined with zeros in corresponding spin-splitting factors $ R_S (2), R_S(3).$ They also depend on the values of $R(1),R(3) $ (which are dependent of temperature and scattering),  and on the magnetic field. Here, we show that the spin-zero positions also depend on the FS geometry at the extremal cross-sections. This follows from the expressions (36), (37).
\begin{figure}[t]
\begin{center}
\includegraphics[width=8.6cm,height=4.3cm]{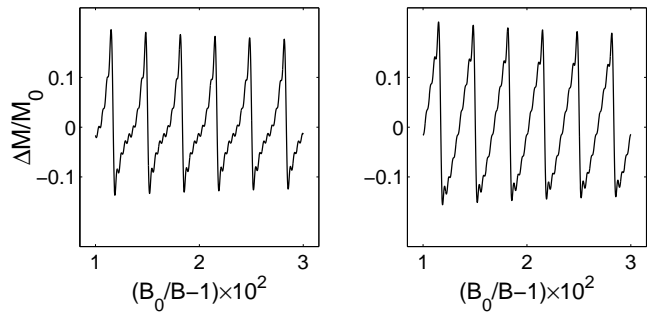}
\caption{The effect of the chemical potential oscillations on the magnetization oscillations. The curves are plotted taking into account the chemical potential oscillations according Eq. (32) (left), and neglecting the latter (right). For both curves $ l=4 \ (\rho = 1.125) $ and the remaining parameters coincide with those used in plotting Fig. 2.
}
\label{rateI}
\end{center}
\end{figure}

\section{v. angular dependence of the oscillations amplitudes}

The effect of the FS curvature on the quantum oscillations in the magnetization is expected to be very sensitive to the geometry of the experiments. The reason is that the effective FS cross-sections (with the minimum/maximum cross-sectional areas) run along the lines of zero curvature (if any) only at certain directions of the magnetic field. When the magnetic field is tilted away from such direction by the angle $\theta, $  the extremal cros-section slips from the nearly cylindrical strip on the FS containing a zero curvature line. This results in the decrease in the oscillations amplitude. The phase of the oscillations also changes.
These angular dependencies of the oscillations amplitudes and phases radically differ in origin from the effect first described by Yamaji \cite{27}. 

The Yamaji effect occurs due to the coincidence of the FS extremal areas $ A_{\max} $ and $ A_{\min} $ at certain angles of inclination of the magnetic field with respect to the FS symmetry axis. At such angles all the cross-sections on the FS have the same area, so the amplitude of the de Haas--van Alphen oscillations increases. The Yamaji effect originates from the periodicity of the $ p_z $ dependent contribution to the charge carriers energy spectrum, and it is inrelated to the presence/absence of zero curvature lines on the relevant FS. Also, there is a crucial difference in the manifestations of the two effects. The angular dependence originating from the effects of the FS curvature reveals itself at the very small values of  $\theta,$ whereas the first maximum due to the Yamaji effect usually appears at $ \theta \sim 10^o $ or even greater. To further clarify the difference between the two effects we analyze the angular dependence of de Haas--van Alphen oscillation amplitudes assuming that the FS curvature becomes zero at an extremal cross-section when the magnetic field is directed along the FS axis of symmetry.   

We suppose that the magnetic field is inclined from the FS symmetry axis by the angle $ \theta $ within the $ xz $ plane, and we employ the coordinate system whose $ z'$ axis is directed along the magnetic field. We use the energy momentum relation given by Eqs. (4), (5), and we rewrite them in terms of new coordinates $ p_z',p_y, p_z'\ \ (p_x'= p_x \cos\theta + p_z \sin\theta;\ \ p_z'= p_z\cos\theta - p_x \sin\theta). $  Changing variables in the Eqs. (4), (5) and keeping in mind that $ \sin \theta$ takes on very small values at small angles $\theta, $ we may present the energy momentum relation for a weakly corrugated Fermi cylinder $(t \ll \mu) $ in the form:
  \bea 
 p_0^2 &=& p_x'^2 \cos^2\theta + p_y^2 - 4m_\perp t
  \nn \\ && \times
 \sum_{n=1}^\infty  \epsilon_n \cos \left (\frac{nd}{\hbar} [p_z'\cos\theta + p_x'\sin \theta]\right)
  \eea
  where $ p_0 $ is the radius of the inwarped Fermi cylinder $(p_0 = \sqrt{A/\pi}).$ Introducing polar coordinates in the cutting plane $(p_x'= p\cos \varphi;\ p_y = p\sin\varphi)$ we may calculate the FS cross-sectional area provided that the magnetic field is tilted away from the FS symmetry axis:
  \be 
 A(p_z',\theta) = \int_0^{2\pi} d \varphi \int_0^p p dp
= A_0 (\theta) + \Delta A (p_z',\theta) .
  \ee
 Here,
  \be 
 A_0 (\theta) = 2 p_0^2 \int_0^{\pi/2} \frac{d\varphi}{\cos^2\theta \cos^2 \varphi + \sin^2\varphi} = \frac{A}{\cos\theta}
  \ee 
  and
  \bea 
 \Delta A (p_z',\cos\theta) &=& 2m_\perp t
\sum_{n=1}^\infty \epsilon_n \int_0^{2\pi} \cos \bigg (
\frac{np_z'd}{\hbar} \cos\theta
  \nn \\  &&+
\frac{np_0 d}{\hbar} \tan \theta \cos\varphi \bigg) d \varphi.
  \eea
Then we can present the oscillating part of the longitudinal magnetization in the form:
  \bea 
 \Delta M_{||} & =& - 2 N \beta \frac{\omega}{\omega_0} \sum_{r=1}^\infty \frac{(-1)^r}{\pi r} 
   \nn \\ && \times
\sin \left(\frac{2\pi rF(\theta)}{B} + 
\Phi_r(\theta)\right)Y_r(\theta).
  \eea
   Here, $ Y_r (\theta) = \sqrt{C^2_r(\theta) + S^2_r (\theta)},$ \\ $ \Phi = \arctan[S_r(\theta)/C_r(\theta)],$ $ F(\theta) = F/\cos\theta,$
  \be 
 C_r(\theta) = \frac{d}{2\pi\hbar\cos\theta}
\int_{(-\pi\hbar/d)\cos\theta}^{(\pi\hbar/d)\cos\theta} \cos\left(\frac{r\lambda^2}{\hbar^2} \Delta A (p_z',\theta)\right) dp_z',
 \ee
 \be 
 S_r(\theta) = \frac{d}{2\pi\hbar\cos\theta}
\int_{(-\pi\hbar/d)\cos\theta}^{(\pi\hbar/d)\cos\theta} \sin\left(\frac{r\lambda^2}{\hbar^2} \Delta A (p_z',\theta)\right) dp_z',
 \ee
 
 Expanding the integrand in (41) in Bessel functions we can easily carry out the integration over $ \varphi.$ Then we get:
  \bea 
 \Delta A (p_z',\cos\theta)& = &
4\pi m_\perp t \sum_{n=1}^\infty 
\epsilon_n \cos\left(\frac{np_z' }{\hbar} \cos \theta\right)
  \nn \\ &&\times
J_0 \left(\frac{np_0 d}{\hbar}\tan\theta\right).
  \eea
  The first term in this expansion coincides with the result given in the Yamaji work. The latter was obtained assuming the simple cosine  warping of the FS described with the Eq. (2).

To analyze the effect of the FS curvature we assume that the curvature becomes zero at $ p_z = \pi\hbar/d. $ Requiring that $ \big(d^2 A/dp_z^2\big)_{p_z=\pi\hbar/d} = 0 $ (see Eq. (3)) and keeping only two first terms in the expansion (5) we obtain:
  \be 
 \epsilon \left(\frac{p_z d}{\hbar}\right) = \cos
\left(\frac{p_z d}{\hbar}\right) + \frac{1}{4} \cos 
\left(\frac{2p_z d}{\hbar}\right)
  \ee
  In this case the cross-sectional area $ A(p_z) $ near $ p_z = \pi\hbar/d $ is approximated as:
  \be 
 A(p_z) = A(p_0) + \frac{\pi m_\perp t}{2}\left (\frac{d}{\hbar}\right)^4 \left(p_z - \frac{\pi\hbar}{d}\right)^4.
  \ee
  Comparing this expression with the Eq. (18) we conclude that the shape parameter $ l = 2.$

Correspondingly, we must put $ \epsilon_1 = 1,$ $ \epsilon_2 = 1/4,$ $\epsilon_n = 0 \ (n > 2) $ into the general expression (46) for $ \Delta A (p_z', \theta). $ As a result we get:
 \bea 
 \Delta A (p_z',\cos\theta)\! &=& \! 4\pi m_\perp t\bigg[\cos \left(\frac{p_z'd}{\hbar} \cos\theta\right) 
J_0  \left(\frac{p_0 d}{\hbar} \tan\theta\right)
  \nn\\\! && + \frac{1}{4}
\cos \left(\frac{2p_z'd}{\hbar} \cos\theta\right) 
J_0  \left(\frac{2p_0 d}{\hbar} \tan\theta\!\!\right) \bigg].\nn\\
  \eea
 To describe FSs possessing closer proximity to a perfect cylinder near the certain extremal cross-section we must keep more terms in the expansion (5). For instance, assuming $ \epsilon_1 = 1, $ $ \epsilon_2 = 2/5,$ $ \epsilon_3 = 1/15, $ and $\epsilon_n = 0 \ (n> 3) $ we ensure that both $ d^2A/d p_z^2 $ and $ d^4A/ dp_z^4 $ become zero at $ p_z = \pi\hbar/d, $ which corresponds to $ l = 3 .$ Similarly, at $ \epsilon_1 = 1, $ $\epsilon_2 = 1/2, $ $ \epsilon_3 = 1/7,\ \epsilon_4 = 1/56, $ and $ \epsilon_n = 0\ (n>4) $ we obtain $ l = 4, $ and so forth. Substituting these numbers into the general expression (45) for $ \Delta A(p_z',\cos\theta), $ and using the results to compute the functions $ C_r(\theta) $ and $ S_r(\theta) $ given by the Eqs. (43), (44), we may finally calculate the factors $ Y_r(\theta).$ The latter describe the desired angular dependence of the oscillations amplitudes.

\begin{figure}[t]
\begin{center}
\includegraphics[width=6.2cm,height=6cm]{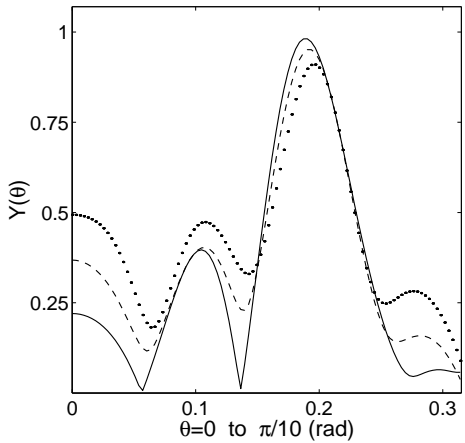}
\caption{Angular dependencies of the magnetization oscillations amplitudes. The curves are plotted assuming that $ t/\hbar\omega = 0.5,\ p_0 d/\hbar = 4\pi.$ The shape parameter $ l $ takes on the values: $l = 1 $ (solid line); $ l = 2$ (dashed line);  $ l = 3 $ (dotted line). The plotted curves are described by the Eqs. (42)--(44) $(r= 1) $.
}
\label{rateI}
\end{center}
\end{figure}

Here, we carried out the calculations accepting $ t/\hbar\omega = 0.5 $ and $ p_0 d/\hbar = 4\pi $ and keeping only the first term in the sum over $ r $ in the Eq. (42). The resulting curves are presented in the Figure 4. The solid line in this figure is associated with the energy spectrum of the form (2).
The corresponding FS has a cosine warping, and possesses a nonzero curvature at the expremal cross-sections. The high peak at $ \theta = 0.185\ (10.6^o) $ corresponds to the first Yamaji maximum. The position of this peak is in agreement with the equation $ (p_0 d/\hbar)\tan\theta = 3\pi/4 $ (see \cite{27}). Two preceding zeros originate from the beats. The remaining curves represent FSs whose curvature becomes zero at their minimum cross-sections at $ \theta = 0. $ We see that the closer the FS shape to a that of a perfect cylinder in the vicinities of these cross-sections (the greater is the value of $l $) the greater is the amplitude of the oscillations near $ \theta = 0 $ and the smaller is the Yamaji maximum. At $ l = 3$  the Yamaji maximum is approximately $2  $ times higher than the maximum at $ \theta = 0, $ whereas at $ l = 1 $  the ratio of the heights takes on the value close to $4.$  We may expect that at very close proximity of the FS to a cylinder near the extremal cross-section $(l \sim 10), $ the amplitude maximum at $ \theta = 0 $ will exceed the Yamaji peak.  Also, we see that the amplitudes of the oscillations associated with the FSs containing zero curvature extremal cross-sections do not become zero at small angles $\theta. $ This is due to the fact that the considered FSs are warped cylinders whose curvature is zero at the cross-sections with the minimum areas (such as $p_z = \pi\hbar/d)$ but remains nonzero at the cross-sections with the maximum areas (such as $p_z = 0).$ 
Difference in the amplitudes of the oscillations originating from the FS local geometry near its extremal cross-sections prevents beats from being well manifested.

The angular dependence of the magnetization oscillations amplitude resembling that presented in the Fig. 4 was reported to be observed in experiments on the Q2D organic metal $\alpha-(BETS)_2 TIHg(SeCN)_4$ \cite{28}. In these experiments a high peak in the amplitude was observed when the magnetic field was directed along the axis of the corrugated cylinder which is the part of the FS. When the field was tilted away from this axis by the angle $ \theta $ the amplitude was rapidly reducing, and it reached approximately one half of the initial value at $ \theta \sim 5^o. $ Further increase in the angle $ \theta $ brought small variations in the amplitude until another peak was reached at $\theta \sim 18^o. $ Identifying this second peak with the first Yamaji maximum we may conjecture that the higher peak at $ \theta = 0 $ arises due to the presence of the FS extremal cross-sections of zero curvature. 
The relation between the heights of the peaks reported in \cite{27} gives grounds to expect that the nearly cylindrical segments of the $\alpha-(BETS)_2TIHg(SeCN)_4 $ Fermi surface (where the FS curvature becomes zero) are very close to perfect cylinders, so that the shape parameter $ l $ takes on values significantly greater than unity.

\section{vi. Conclusion}
In summary, the present work  aims to study the effect of the FS shape on the de Haas-van Alphen oscillations in Q2D conductors. Such analysis is important for there exists a great deal of interest in studies of band-structure parameters and other electronic properties of these materials. Usually, the simple model for the electron spectrum (2) is employed to extract the relevant information from the experiments. This approximation has its limitations, so some problems arise in interpreting the experimental data (see e.g. Refs. \cite{15,16,28}). An important limitation of the current theory is that the latter misses the effects of the FS geometry assuming a simple cosine warping of the FS. Here, we lift this restriction on the FS shape. We show that the FS profile may significantly affect the quantum oscillations in magnetization if the FS curvature becomes zero at a cross-sectional area. Also, we show that main characteristics of the oscillations are determined by two different factors, namely by the FS curvature at the effective cross-sections and the transfer integral, whereas existing theory takes into account only the latter. These two factors work simultaneously, and their effects could be separated. As discussed above the features typical for 2D conductors could be revealed when the FS is rippled provided that its curvature turns zero at the cross-sections with extremal areas. 

The presence of zero curvature effective cross-sections noticeably affects the angular dependencies of the oscillation amplitudes. A maximum originating due to the FS local geometry could emerge. The height of this peak could be comparable with the height of the well known Yamaji maximum and even exceed the latter. This agrees with the experimental results observed on the Q2D $\alpha-(BETS)_2TIHg(SeCH)_4$ organic metal \cite{28}.
The proposed approach could be useful in analyzing experiments on magnetization oscillations in Q2D conductors, especially those where sawtooth features in the oscillations are well pronounced. It could help to extract important extra informations concerning fine geometrical features of the FSs of such materials.

\section{ Acknowledgments}
The author thanks G. M. Zimbovsky for help with the manuscript. This work was supported  by NSF Advance program SBE-0123654 and PR Space Grant NGTS/40091.

\end{document}